\documentclass[12pt]{article}
\usepackage{eurosym}
\usepackage{amssymb}
\usepackage{amsmath}
\usepackage{cite}
\usepackage{lscape}
\usepackage{graphicx}
\usepackage{epsf}

\newtheorem{theorem}{Theorem}

\newtheorem{corollary}[theorem]{Corollary}

\newtheorem{proposition}[theorem]{Proposition}

\input{tcilatex}
\begin{document}

\title{Lie symmetry classification for the 1+1 and 1+2 generalized Zoomeron
equations}
\author{Andronikos Paliathanasis$^{1,2}$\thanks{%
Email: anpaliat@phys.uoa.gr} and P.G.L Leach$^{1}$ \\
{\ }$^{1}${\textit{Institute of Systems Science, Durban University of
Technology }}\\
{\ \textit{PO Box 1334, Durban 4000, Republic of South Africa}\ } \\
$^{2}${\textit{Departamento de Matem\'{a}ticas, Universidad Cat\'{o}lica del
Norte,}}\\
{\ \textit{Avda. Angamos 0610, Casilla 1280 Antofagasta, Chile}}}
\maketitle

\begin{abstract}
We present a complete algebraic classification of the Lie symmetries for
generalized Zoomeron equations. For the generalized 1+1 and 2+1 Zoomeron
equations we solve the Lie symmetry conditions in order to constrain the
free functions of the equations. We find that the differential equations of
our consideration admit the same number of Lie symmetries with the
non-generalized equations. The admitted Lie symmetries form the Lie algebras 
$2A_{1}$, $A_{3,3}$ $\ $for the 1+1 generalized Zoomeron equation, and the $%
A_{4,5}^{ab}\,$, $3A_{1}\otimes _{s}2A_{1}$ in the case of the 2+1
generalized Zoomeron equation. The one-dimensional optimal system is
constructed for the two equations and similarity solutions are derived.\ The
similarity transformation lead to the derivation of kink solutions. Indeed,
the similarity exact solutions determined in this work are asymptotic
solutions near to the singular behaviour of the kink behaviour.

\bigskip

\end{abstract}

\section{Introduction}

\label{sec1}

Nonlinear differential equations play a significant role in applied science
because they are the main approach to the mathematical description of real
world phenomena \cite{bok1}. The Korteweg-de Vries (KdV) equation is
probably the most fully understood and well studied 1+1 nonlinear partial
differential equation which arises in various physical phenomena. Indeed,
the KdV equation provides exact solutions which are used for the
mathematical description of waves in various media, such as surface gravity
waves, internal waves in the ocean, acoustic waves, optical waves and many
others \cite{kdv1,kdv2,kdv3}.

With the study of KdV the concept of solitons was introduced. Solitons are
solitary wave solutions the shape and speed of which are invariant after the
collision, that is, the solitary waves have particle-like properties. The
first known soliton was found in the Fermi-Pasta-Ulam lattice \cite{sf1}.
There are many studies in the literature where the KdV is modified so as to
describe physical phenomena. For more details we refer the reader to \cite%
{sf2}. Moreover, the concept of topological solitons has been also
introduced \cite{sf3}. Topological solitons are exact solutions in classical
field theory and they describe particle-like objects such as the Yang-Mills
instatons or the Skyrmions \cite{sf4,sf5,sf6}.

In the attempt to generalize the KdV equation there has been introduced an
1+1 nonlinear partial differential equation known as the Zoomeron equation 
\cite{zm00}%
\begin{equation}
\left( \frac{u_{xt}}{u}\right) _{tt}-\left( \frac{u_{xt}}{u}\right)
_{xx}+2\left( u^{2}\right) _{xt}=0.  \label{sz.01}
\end{equation}%
The dynamics, the analytic solutions and the algebraic properties of
equation (\ref{sz.01}) have been widely studied in the literature, see for
instance \cite{zm01,zm02} and references therein.\ 

The novelty of equation (\ref{sz.01}) is that describes solitons which move
with a variable speed and there exist nontrivial polarization effects. These
solitons are categorized in two families; the accelerated solitons which are
coming in the past from one side and boomeranging back to that side with the
same speed in the remote future, the second family of solutions have the
characteristic to describe trapped oscillating in a spatial domain changing
direction multiple times around a fixed point; that is,. a potential well.
The latter solutions are known as boomeron and trappon solutions \cite{zm00}%
. 

The 2+1 Zoomeron equation, 
\begin{equation}
\left( \frac{u_{xy}}{u}\right) _{tt}-\left( \frac{u_{xy}}{u}\right)
_{xx}+2\left( u^{2}\right) _{xt}=0,  \label{sz.02}
\end{equation}%
was introduced in \cite{sm03} in which the $\left( \frac{G^{\prime }}{G}%
\right) -$expansion method was applied to find exact solutions.\
Furthermore, travelling wave solutions for equation (\ref{sz.02}) were found
by using the MSE method and the Exp-function method in \cite{sm03a,sm03b}.
The Lie symmetries of equation (\ref{sz.02}) were the subject of study in 
\cite{sm04}.\ It was found that equation (\ref{sz.02}) admits a
five-dimensional Lie algebra and the Lie symmetries were applied to
determine new similarity solutions. 

Equation (\ref{sz.02}) is of special interest because it describes specific
solitons with important features related to the nonlinear optics and fluid
mechanics \cite{sm0aA}. In \cite{sm1aA} the physical properties of the
solitons provided by equation (\ref{sz.02}) investigated in details. The
authors determined solutions described by rational, trigonometric and
hyperbolic functions; consequently, kink, singular solitons, periodic
solitons and lump-like solitons can be described by equation (\ref{sz.02}).
The authors discussed how equation (\ref{sz.02}) can be used to describe
foam drainage phenomena \cite{sm2aA}. Solitary wave solutions derived in 
\cite{sm3aA} with the application of the tanh-coth approach, while new
periodic solutions determined in \cite{sm4aA} by the use of the sine-cosine
method. For more 

A generalization of equation (\ref{sz.02}) was introduced in \cite{sm05} 
\begin{equation}
\left( \frac{u_{xy}}{u}\right) _{tt}-k^{2}\left( \frac{u_{xy}}{u}\right)
_{xx}+2\alpha \left( u^{2n}\right) _{xt}=0.  \label{sz.03}
\end{equation}%
For the latter 2+1 generalized equation it was found that it admits a
five-dimensional Lie algebra which forms the $3A_{1}\otimes _{s}2A_{1}$ Lie
algebra for$~n>0$~\cite{patera}. With the use of Lie symmetry analysis exact
solutions were found. Specifically, equation (\ref{sz.03}) under the
application of similarity transformations can be written in the form of the
Airy equation, the Lane-Emden equation or the Bernoulli equation.
Consequently, trappon and boomeron solutions are provided by equation (\ref%
{sz.03}).

Lie symmetry analysis is a systematic approach for the study of nonlinear
differential equations \cite{ibra,Bluman,Stephani,olver}. The novelty of the
Lie symmetry approach is that though a systematic approach the existence of
invariant transformations can be determined. The latter can be used to
simplify the given differential equation with the use of similarity
transformations. Under the application of similarity transformations in a
given differential equation the number of independent variables can be
reduced or the order of the differential equation can be reduced.
Furthermore, conservation laws can be constructed which are essential for
the study of the properties for the given differential equation \cite{ibra}.
Lie symmetries have been applied for the study of nonlinear differential
equations in all areas of applied mathematics \cite{swr06,ans2d,m7,m8,m9,r2}%
. In \cite{sol1}, the Lie symmetries were used for the study of the
nonlinear Schr\"{o}dinger equation and for the construction of exact
solutions which describe solitons.\ Furthermore, solitons of the 2+1
dimensional Nizhnik-Novikov-Vesselov equation were determined by using the
Lie symmetry analysis in \cite{sol2}.

The Lie symmetries for the 2+1 generalized Hirota-Satsuma-Ito equation
investigated in \cite{sol3} where new conservation laws and new soliton
solutions were determined.\ Applications of the symmetry analysis in
higher-dimensional partial differential equations or in higher-order
equations are presented for instance in \cite{sol4,sol5,sol6,sol7,sol8} and
references therein. Specifically, in \cite{sol7} the Lie symmetries for
multidimensional difference equations investigate, in the case of a
two-dimensional lattice. Moreover, in \cite{sol8} the authors applied the
Lie symmetries for the study of multi-dimensional parabolic partial
differential equations with diffusion components. 

In this work we are interested in the extension of the analysis performed in 
\cite{sm05}. Indeed we perform a complete classification of the Lie
symmetries for the generalized 1+1 $\left( gZE^{1}\right) $ and 2+1 $\left(
gZE^{2}\right) $ Zoomeron equations of the form%
\begin{equation}
gZE^{1}:\left( \frac{u_{xt}}{f\left( u\right) }\right) _{tt}-\left( \frac{%
u_{xt}}{f\left( u\right) }\right) _{xx}+2\left( g\left( u\right) \right)
_{xt}=0,  \label{sz.04}
\end{equation}%
\begin{equation}
gZE^{2}:\left( \frac{u_{xy}}{f\left( u\right) }\right) _{tt}-\left( \frac{%
u_{xy}}{f\left( u\right) }\right) _{xx}+2\left( g\left( u\right) \right)
_{xt}=0,  \label{sz.05}
\end{equation}%
where $f\left( u\right) $,$~g\left( u\right) \,\ $are nonconstant functions,
that is, $f_{,u}g_{,u}\neq 0$. In particular, inspired by the Ovsiannikov
classification scheme \cite{Ovsi} we determine the unknown functions $%
f\left( u\right) $,$~g\left( u\right) $ such that equations (\ref{sz.04})
and (\ref{sz.05}) to admit Lie symmetries.

Such an analysis will provide us with important information about the
relation of the Lie symmetries with the existence of solitons. Furthermore,
from the Lie symmetries we can construct similarity transformations and
determined exact solutions for the latter equations. These exact solutions
are mainly related with the existence of travel-wave solutions or
self-similar solutions. Hence, the exact solutions provide the asymptotic
behaviour of the general solution near to the moving wall as the soliton
\textquotedblleft moves\textquotedblright\ in the space or time variables.
The plan of the paper is as follows.

The basic properties and definitions for the Lie symmetry analysis are
presented in Section \ref{sec2}. In\ Section \ref{sec3} we present the Lie
symmetry classification for the generalized 1+1 and 2+1 Zoomeron equations.
The application of the Lie symmetries is presented in Section \ref{sec4} in
which we use the similarity transformations in order to construct exact
solutions. Finally, in Section \ref{sec5} we summarize our results and we
draw our conclusions.

\section{Preliminaries}

\label{sec2}

In this Section we present the basic properties and definitions for the
theory of Lie symmetries of differential equations. Assume the function $%
\Phi $ which describes the map of a one-parameter point transformation such
as $\Phi \left( \mathbf{u}\left( t,x^{i}\right) \right) =\mathbf{u}\left(
t,x^{i}\right) \ $with infinitesimal transformation%
\begin{eqnarray}
t^{\prime } &=&t^{i}+\varepsilon \xi ^{t}\left( t,x^{i},u\right) ,
\label{sv.12} \\
x^{\prime } &=&x^{i}+\varepsilon \xi ^{x^{i}}\left( t,x^{i},u\right) , \\
u^{\prime } &=&u+\varepsilon \eta \left( t,x^{i},u\right) ,  \label{sv.13}
\end{eqnarray}%
and generator $\mathbf{X}=\frac{\partial t^{\prime }}{\partial \varepsilon }%
\partial _{t}+\frac{\partial x^{\prime }}{\partial \varepsilon }\partial
_{x}+\frac{\partial u}{\partial \varepsilon }\partial _{u},~$where~{$%
\varepsilon $ is the parameter of smallness}; $u\left( t,x^{i}\right) \ $is
the dependent function and $\left( t,x^{i}\right) $ are the independent
variables with $x^{i}=\left( x,y\right) $.

Let $u\left( t,x^{i}\right) $ be a solution to the differential equation $%
\mathcal{H}\left( t,x,u,u_{,t},u_{,x},...\right) =0.$ Then, under the
one-parameter map $\Phi $, function $u^{\prime }\left( t^{\prime
},x^{i\prime }\right) =\Phi \left( u\left( t,x^{i}\right) \right) $ is a
solution for the differential equation, $\mathcal{H}$, if and only if the
differential equation is also invariant under the action of the map, $\Phi $%
, that is, the following condition holds \cite{ibra,Bluman,Stephani,olver} 
\begin{equation}
\Phi \left( \mathcal{H}\left( t,x,u,u_{t},u_{x}...\right) \right) =0.
\label{f1}
\end{equation}

For every map $\Phi $ in which condition (\ref{f1}) holds, it means that the
generator~$X$ is a Lie point symmetry for the differential equation, while
the following condition is true 
\begin{equation}
\mathbf{X}^{\left[ n\right] }\left( \mathcal{H}\left(
t,x,u,u_{t},u_{x}...\right) \right) =0.  \label{sv.17}
\end{equation}%
Vector field $\mathbf{X}^{\left[ n\right] }$ describes the first extension
of the symmetry vector in the jet-space of variables, $\left\{
t,x,u,u_{t},u_{x},....,u_{i_{1},...,i_{n}}\right\} $.

The importance of the existence of a Lie symmetry for a given differential
equation is that from the associated Lagrange's system,%
\begin{equation}
\frac{dt}{\xi ^{t}}=\frac{dx^{i}}{\xi ^{x^{i}}}=\frac{du}{\eta },
\end{equation}%
the solution of this system provides the invariant functions which can be
used to reduce the number of the independent variables of the differential
equation and lead to the construction of similarity solutions In the case of
partial differential equations, the application of the Lie invariants
reduces the number of the independent variables.

The admitted symmetry vectors of a given set of differential equations
constitute a closed-group known as a Lie group. \ The main application of
the Lie symmetries is the determination of solutions known as similarity
solutions and follows from the application of the Lie invariants in the
differential equations. However, in order to classify all the possible
similarity transformations and solutions the one-dimensional optimal system
should be calculated \cite{olver}.

Assume the $n$-dimensional Lie algebra $G_{n}$ with elements $\left\{
X_{1},~X_{2},~...~X_{n}\right\} ~$and structure constants $C_{jk}^{i}$. We
define the two symmetry vectors 
\begin{equation}
Z=\sum_{i=1}^{n}a_{i}X_{i}~,~W=\sum_{i=1}^{n}b_{i}X_{i}~,~\text{\ }%
a_{i},~b_{i}\text{ are constants.}
\end{equation}%
and we define the Adjoint operator 
\begin{equation}
Ad\left( \exp \left( \epsilon X_{i}\right) \right) X_{j}=X_{j}-\epsilon 
\left[ X_{i},X_{j}\right] +\frac{1}{2}\epsilon ^{2}\left[ X_{i},\left[
X_{i},X_{j}\right] \right] +...~,  \label{sw.07}
\end{equation}%
known as the adjoint representation, in which $\left[ X_{i},X_{j}\right]
=C_{ij}^{k}X_{k}$ is the Lie Bracket.

We say that the vectors $Z$ and $W$ are equivalent if and only if \cite%
{olver} 
\begin{equation}
\mathbf{W}=\prod_{i=1}^{n}Ad\left( \exp \left( \epsilon _{i}X_{i}\right)
\right) \mathbf{Z,}
\end{equation}%
or 
\begin{equation}
W=cZ~,~c=const\text{ that is }b_{i}=ca_{i}\text{.}
\end{equation}

As far as the constants $a_{i}$ are concerned, their independent constants
are invariant functions of the Adjoint operator. Indeed if $\phi \left(
a_{i}\right) $ is a function of $a_{i}$, the invariants are given by the
linear system of partial differential equations 
\begin{equation}
\Delta _{i}\left( \phi \right) =C_{ij}^{k}a^{j}\frac{\partial \phi }{%
\partial a^{i}}\equiv 0.  \label{hm.19}
\end{equation}%
The one-dimensional subalgebras of $G_{n}$ which are not related through the
adjoint representation form the one-dimensional optimal system. The
determination of the one-dimensional system is essential in order to perform
a complete classification of all the possible similarity transformations and
solutions.$~$

\section{Lie symmetries for the Zoomeron equations}

\label{sec3}

We apply the Lie symmetry condition (\ref{sv.17}) for the generalized 1+1 $%
\left( gZE^{1}\right) $ and 2+1 $\left( gZE^{2}\right) $ Zoomeron equations
and we determine all the functional forms of $f\left( u\right) $ and $%
g\left( u\right) $ where the partial differential equations admit Lie point
symmetries. For simplicity on our presentation we omit the calculations. The
results are summarized in the following propositions.

For the $gZE^{1}$ equation it follows.

\begin{proposition}
The Lie point symmetry classification for the generalized 1+1 Zoomeron
equation (\ref{sz.04}) for arbitrary functions $f\left( u\right) $ and $%
g\left( u\right) $ are two, they are $X_{1}=\partial _{t}$ and $%
X_{2}=\partial _{x}$ with the commutator, $\left[ X_{1},X_{2}\right] =0$,
that is, they form the $2A_{1}$ Lie algebra. However for specific functional
forms of $f\left( u\right) $ and $g\left( u\right) $ additional Lie
symmetries exist.

(A) For $f\left( u\right) =f_{0}u^{K}$ and $g\left( u\right) =g_{0}+g_{1}u^{-%
\frac{2}{a_{1}}-\left( K-1\right) }$, where $\alpha _{1}$ is a nonzero
constant; in this case $gZE^{1}$ equation admits the additional Lie symmetry 
$X_{3}=t\partial _{t}+x\partial _{x}+\alpha _{1}u\partial _{u}$. The nonzero
commutators for the Lie algebra are $G_{3}=\left\{ X_{1},X_{2},X_{3}\right\} 
$ are $\left[ X_{1},X_{3}\right] =X_{1}$ and $\left[ X_{2},X_{3}\right]
=X_{2}$. Thus, $G_{3}$ is the $A_{3,3}$ Lie algebra.

(B) For $f\left( u\right) =f_{0}u^{K}$ and $g\left( u\right) =g_{0}+g_{1}\ln
u$, $gZE^{1}$ equation admits the additional Lie symmetry$~X_{4}=t\partial
_{t}+x\partial _{x}+\frac{1}{1-K}u\partial _{u}$, which, as above, the
admitted Lie symmetry vectors $\left\{ X_{1},X_{2},X_{4}\right\} $ form the $%
A_{3,3}$ Lie algebra.

(C) For $f\left( u\right) =f_{0}\exp \left( Ku\right) $ and $g\left(
u\right) =g_{0}+g_{1}\exp \left( -\left( K+\frac{2}{\alpha _{2}}\right)
u\right) $, with $\alpha _{2}$ a nonzero constant; the additional Lie
symmetry vector for equation $gZE^{1}$ is the field $X_{5}=t\partial
_{t}+x\partial _{x}+a_{2}\partial _{u}$. The Lie symmetries $\left\{
X_{1},X_{2},X_{5}\right\} $ form the $A_{3,3}$ Lie algebra.

(D) For $f\left( u\right) =f_{0}\exp \left( Ku\right) $ and $g\left(
u\right) =g_{0}+g_{1}u$ the additional Lie symmetry vector is the vector
field $X_{6}=t\partial _{t}+x\partial _{x}-\frac{1}{K}\partial _{u}$. The
admitted Lie symmetries form the $A_{3,3}$ Lie algebra.
\end{proposition}

Similarly for the $gZE^{2}$ equation the Lie symmetry classification is as
follows.

\begin{proposition}
The generalized 2+1 Zoomeron equation (\ref{sz.05}) for arbitrary functional
forms of $f\left( u\right) $ and $g\left( u\right) $ is invariant under the
four-dimensional Lie algebra $G_{4}=\left\{ Y_{1},Y_{2},Y_{3},Y_{4}\right\} $
which consists of the vector fields $Y_{1}=\partial _{t}$, $Y_{2}=\partial
_{x}$,~$Y_{3}=\partial _{y}$ and $Y_{4}=t\partial _{t}+x\partial
_{x}-y\partial _{y}$. The nonzero commutators for the elements of the $G_{4}$
are $\left[ Y_{1},Y_{4}\right] =Y_{1}$, $\left[ Y_{2},Y_{4}\right] =Y_{2}$
and $\left[ Y_{3},Y_{4}\right] =-Y_{3}$. Hence, it follows that $G_{4}$
forms the $A_{4,5}^{ab}$ Lie algebra. For specific forms of $f\left(
u\right) $ and $g\left( u\right) $ additional symmetries exist.\newline
(I) When $f\left( u\right) =f_{0}u^{K}$ and $g\left( u\right)
=g_{0}+g_{1}u^{-\frac{1}{\alpha _{1}}-\left( K-1\right) }$, where $\alpha
_{1}$ is a nonzero constant; in this case $gZE^{2}$ equation admits the
additional Lie symmetry vector $Y_{5}=y\partial _{y}-\alpha _{1}u\partial
_{u}$. The additional nonzero commutator is the $\left[ Y_{3},Y_{5}\right]
=Y_{3}$. Thus, the Lie algebra $G_{5}=\left\{
Y_{1},Y_{2},Y_{3},Y_{4},Y_{5}\right\} $ is the $3A_{1}\otimes _{s}2A_{1}$.

(II) For $f\left( u\right) =f_{0}u^{K}$ and\thinspace\ $g\left( u\right)
=g_{0}+g_{1}\ln \left( u\right) $, the generalized $gZE^{2}$ equation admits
the additional Lie symmetry vector $Y_{6}=y\partial _{y}+\frac{1}{1-K}%
u\partial _{u}$. The admitted Lie symmetries form the $3A_{1}\otimes
_{s}2A_{1}$ Lie algebra.

(III) Moreover, when $f\left( u\right) =f_{0}\exp \left( Ku\right) $ and $%
g\left( u\right) =g_{0}+g_{1}\exp \left( -\left( K+\frac{2}{\alpha _{2}}%
\right) u\right) $ the additional Lie symmetry vector for equation $gZE^{2}$
is the vector field~$Y_{7}=y\partial _{y}+a_{2}\partial _{u}$, where again
the admitted Lie algebra is the $3A_{1}\otimes _{s}2A_{1}$.

(IV) Finally, for $f\left( u\right) =f_{0}\exp \left( Ku\right) $ and $%
g\left( u\right) =g_{0}+g_{1}u$, equation $gZE^{2}$ is invariant under the
action of the one-parameter point transformation with generator the vector
field $Y_{8}=y\partial _{y}-\frac{1}{K}\partial _{u}$. Similarly, the
admitted Lie symmetries form the $3A_{1}\otimes _{s}2A_{1}$ Lie algebra.
\end{proposition}

As far as the admitted Lie algebras by the generalized Zoomeron equations
the following corollary holds.

\begin{corollary}
The generalized 1+1 Zoomeron equation (\ref{sz.04}) is invariant under the
action of a two-dimensional or three-dimensional Lie algebra, the $2A_{1}$
or the $A_{3,3}$. Furthermore, the generalized 2+1 Zoomeron equation (\ref%
{sz.05}) admits four or five Lie symmetries which form the $A_{4,5}^{ab}$ or
the $3A_{1}\otimes _{s}2A_{1}$ Lie algebras.
\end{corollary}

At this point it is important to mention that not all the comments $f_{0}$, $%
g_{0}$ and $g_{1}$ are essential parameters. Without loss of generality in
the following we assume $g_{0}=0$ and $f_{0}=1$. We proceed with the
derivation of the one-dimensional optimal system for each of the admitted
Lie algebras.

\subsection{One-dimensional optimal system}

\label{sec3.1}

The main application of Lie symmetries is the definition of similarity
transformations which are used to simplify the given differential equation.
The derivation of the one-dimensional optimal system is essential for the
complete classification of all the independent similarity transformations,
consequently, of the independent similarity solutions.

In Tables \ref{tab01}, \ref{tab02} and in Tables \ref{tab1}, \ref{tab2} we
present the commutators and the Adjoint representation for the admitted Lie
algebras of equations $gZE^{1}$ and $gZE^{2}$ respectively.

\begin{table}[tbp] \centering%
\caption{Commutators of the admitted Lie point symmetries for the
generalized 1+1 Zoomeron equation, $A=3,4,5,6$}%
\begin{tabular}{cccc}
\hline\hline
$\left[ ~,~\right] $ & $\mathbf{X}_{1}$ & $\mathbf{X}_{2}$ & $\mathbf{X}_{A}$
\\ 
$\mathbf{X}_{1}$ & $0$ & $0$ & $X_{1}$ \\ 
$\mathbf{X}_{2}$ & $0$ & $0$ & $X_{2}$ \\ 
$\mathbf{X}_{A}$ & $-X_{1}$ & $-X_{2}$ & $0$ \\ \hline\hline
\end{tabular}%
\label{tab01}%
\end{table}%

\begin{table}[tbp] \centering%
\caption{Adjoint representation for the admitted Lie point symmetries of the
generalized 1+1 Zoomeron equation, $A=3,4,5,6$}%
\begin{tabular}{cccc}
\hline\hline
$Ad\left( \exp \left( \varepsilon \mathbf{X}_{\mu }\right) \right) \mathbf{X}%
_{\nu }$ & $\mathbf{X}_{1}$ & $\mathbf{X}_{2}$ & $\mathbf{X}_{3}$ \\ 
$\mathbf{X}_{1}$ & $X_{1}$ & $X_{2}$ & $X_{3}-\varepsilon X_{1}$ \\ 
$\mathbf{X}_{2}$ & $X_{1}$ & $X_{2}$ & $X_{3}-\varepsilon X_{2}$ \\ 
$\mathbf{X}_{3}$ & $e^{\varepsilon }X_{1}$ & $e^{\varepsilon }X_{2}$ & $%
X_{3} $ \\ \hline\hline
\end{tabular}%
\label{tab02}%
\end{table}%

We assume now that the generic symmetry vector is 
\begin{equation}
X=a_{1}X_{1}+a_{2}X_{2}+a_{A}X_{A},
\end{equation}%
for equation $gZE^{1}$, or 
\begin{equation}
Y=b_{1}Y_{1}+b_{2}Y_{2}+b_{3}Y_{3}+b_{4}Y_{4}+b_{A}Y_{A},
\end{equation}%
for the generalized $gZE^{2}$ equation, where $a_{1},~a_{2},~a_{3},~a_{A}$
and $b_{1},~b_{2}$, $b_{3},~b_{4}$ and $b_{A}$ are constants.

For the admitted two-dimensional Lie algebra $2A_{1}\,\ $of equation $%
gZE^{1},$ with $a_{A}=0$, we determine the invariants of the Adjoint
representation $a_{1}$ and $a_{2}$. Thus, the one-dimensional optimal system
consists of the one-dimensional Lie algebras%
\begin{equation}
OpS\left( 2A_{1}\right) :\left\{ X_{1}\right\} ~,~\left\{ X_{2}\right\}
~,~\left\{ X_{1}+\alpha X_{2}\right\} .
\end{equation}

In addition, for the $A_{3,3}$ Lie algebra we determine that $a_{A}$ is the
invariant for the Adjoint representation. Hence, the one-dimensional optimal
system consists of the vector fields%
\begin{equation}
OpS\left( A_{3,3}\right) :OpS\left( 2A_{1}\right) ~,~\left\{ X_{A}\right\} .
\end{equation}

Similarly, for equation $gZE^{2}$ and the four-dimensional Lie algebra $%
A_{4,5}^{ab}$ we derive the Lie invariants of the Adjoint representation to
be $a_{4}$. The one-dimensional optimal system is%
\begin{eqnarray}
OpS\left( A_{4,5}^{ab}\right) &:&\left\{ Y_{1}\right\} ~,~\left\{
Y_{2}\right\} ~,~\left\{ Y_{3}\right\} ~,~\left\{ Y_{4}\right\} ~,~\left\{
Y_{1}+\alpha Y_{2}\right\}  \notag \\
&&\left\{ Y_{1}+\alpha Y_{3}\right\} ~,~\left\{ Y_{2}+\alpha Y_{3}\right\}
~,~\left\{ Y_{1}+\alpha Y_{2}+\beta Y_{3}\right\} .
\end{eqnarray}

Finally, for the five-dimensional Lie algebra $3A_{1}\otimes _{s}2A_{1}$ $\ $%
the invariants of the Adjoint operator are $a_{4}$ and $a_{5}$. Hence, with
the use of Table \ref{tab2} we derive the one-dimensional system%
\begin{eqnarray}
OpS\left( 3A_{1}\otimes _{s}2A_{1}\right) &:&OpS\left( A_{4,5}^{ab}\right)
~,~\left\{ Y_{A}\right\} ~,~\left\{ Y_{4}+\alpha Y_{A}\right\} ~,~~  \notag
\\
&&\left\{ Y_{1}+\alpha Y_{A}\right\} ~,~\left\{ Y_{2}+\alpha Y_{A}\right\}
~,~\left\{ Y_{1}+\alpha Y_{2}+\beta Y_{A}\right\} .
\end{eqnarray}

\begin{table}[tbp] \centering%
\caption{Commutators of the admitted Lie point symmetries for the
generalized 2+1 Zoomeron equation, $A=5,6,7,8$}%
\begin{tabular}{cccccc}
\hline\hline
$\left[ ~,~\right] $ & \textbf{$Y$}$_{1}$ & \textbf{$Y$}$_{2}$ & \textbf{$Y$}%
$_{3}$ & \textbf{$Y$}$_{4}$ & \textbf{$Y$}$_{A}$ \\ 
\textbf{$Y$}$_{1}$ & $0$ & $0$ & $0$ & $Y_{1}$ & $0$ \\ 
\textbf{$Y$}$_{2}$ & $0$ & $0$ & $0$ & $Y_{2}$ & $0$ \\ 
\textbf{$Y$}$_{3}$ & $0$ & $0$ & $0$ & $Y_{3}$ & $-Y_{3}$ \\ 
\textbf{$Y$}$_{4}$ & $-Y_{1}$ & $-Y_{2}$ & $-Y_{3}$ & $0$ & $0$ \\ 
\textbf{$Y$}$_{A}$ & $0$ & $0$ & $Y_{3}$ & $0$ & $0$ \\ \hline\hline
\end{tabular}%
\label{tab1}%
\end{table}%

\begin{table}[tbp] \centering%
\caption{Adjoint representation for the admitted Lie point symmetries of the
generalized 2+1 Zoomeron equation, $A=5,6,7,8$}%
\begin{tabular}{cccccc}
\hline\hline
$Ad\left( \exp \left( \varepsilon \mathbf{Y}_{\mu }\right) \right) \mathbf{Y}%
_{\nu }$ & \textbf{$Y$}$_{1}$ & \textbf{$Y$}$_{2}$ & \textbf{$Y$}$_{3}$ & 
\textbf{$Y$}$_{4}$ & \textbf{$Y$}$_{A}$ \\ 
\textbf{$Y$}$_{1}$ & $Y_{1}$ & $Y_{2}$ & $Y_{3}$ & $Y_{4}-\varepsilon Y_{1}$
& $Y_{5}$ \\ 
\textbf{$Y$}$_{2}$ & $Y_{1}$ & $Y_{2}$ & $Y_{3}$ & $Y_{4}-\varepsilon Y_{2}$
& $Y_{5}$ \\ 
\textbf{$Y$}$_{3}$ & $e^{\varepsilon }Y_{1}$ & $e^{\varepsilon }Y_{2}$ & $%
Y_{3}$ & $Y_{4}-\varepsilon Y_{3}$ & $Y_{5}+\varepsilon Y_{3}$ \\ 
\textbf{$Y$}$_{4}$ & $e^{\varepsilon }Y_{1}$ & $e^{\varepsilon }Y_{2}$ & $%
e^{\varepsilon }Y_{3}$ & $Y_{4}$ & $Y_{5}$ \\ 
\textbf{$Y_{A}$} & $Y_{1}$ & $Y_{2}$ & $e^{\varepsilon }Y_{3}$ & $Y_{4}$ & $%
Y_{5}$ \\ \hline\hline
\end{tabular}%
\label{tab2}%
\end{table}%

\section{Application of Lie invariants}

\label{sec4}

We proceed with the application of the Lie symmetries to the derivation of
similarity solutions.

\subsection{Equation $gZE^{1}$}

For each element of the one-dimensional optimal system it follows.

\subsubsection{$\left\{ X_{1}\right\} ,~\left\{ X_{2}\right\} $}

The vector fields $\left\{ X_{1}\right\} ~,~\left\{ X_{2}\right\} $ provide
static and stationary solutions of the form $u=u\left( x\right) $ and $%
u=u\left( t\right) $ respectively. These are not solutions of special
interest.

\subsubsection{ $\left\{ X_{1}+\protect\alpha X_{2}\right\} $}

The similarity transformation provided by $\left\{ X_{1}+\alpha
X_{2}\right\} $ is $\xi =x-\alpha t$, $\ u=u\left( \xi \right) $. Thus, by
replacing in (\ref{sz.04}) we find the reduced equation 
\begin{equation}
\left( a^{2}-1\right) \left( \frac{u_{\xi \xi }}{f\left( u\right) }\right)
_{\xi \xi }+2\left( g\left( u\right) \right) _{\xi \xi }=0.
\end{equation}%
For $\alpha ^{2}=1$ it follows $g\left( u\right) =u_{1}\xi +u_{0}$. However,
for $\alpha ^{2}\neq 1$ we find%
\begin{equation}
u_{\xi \xi }+\left( 2\left( g\left( u\right) \right) +u_{1}\xi +u_{0}\right)
f\left( u\right) =0.  \label{fd1}
\end{equation}%
This is a second-order ordinary differential equation of the form $u_{\xi
\xi }+G\left( u\right) +f\left( u\right) \left( u_{1}\xi +u_{0}\right) =0$.
Such equations have been widely studied in the literature before.

When $u_{1}=0$, the analytic solution is expressed in terms of quadratures,
that is,

\begin{equation}
\frac{1}{2}\left( u_{\xi }\right) ^{2}-\int \left( G\left( u\right) +f\left(
u\right) u_{0}\right) du=h~,~h=const\text{.}
\end{equation}

In the case where $g\left( u\right) =u^{2}$, $f\left( u\right) =u$ and $%
u_{1}=0$, $u_{0}=0$ we find the oscillator~$\frac{1}{2}\left( u_{\xi
}\right) ^{2}-\frac{1}{2}u^{2}=h$.

\subsubsection{$\left\{ X_{3}\right\} $}

The Lie symmetry vector $X_{3}$ provides the similarity transformation $%
u=U\left( \sigma \right) t^{a_{1}}$, with $\sigma =\frac{x}{t}$.

In the new variables we derive the closed-form solution of the latter system
is $U\left( \sigma \right) =U_{0}\sigma ^{a_{1}}$, from which we conclude
that 
\begin{equation}
u\left( t,x\right) =u_{0}x^{a_{1}}t^{2a_{1}}.
\end{equation}%
where $u_{0}$ is an integration constant.

\subsubsection{$\left\{ X_{4}\right\} $}

For case (B) of Proposition 1, from the vector field $X_{4}$ we find the
similarity transformation $u=U\left( \sigma \right) t^{\frac{1}{1-K}}$, $%
\sigma =\frac{x}{t}$. We find that in the new variables the closed-form
solution is 
\begin{equation}
u\left( t,x\right) =u_{0}x^{\frac{1}{1-K}}t^{\frac{2}{1-K}}\text{.}
\end{equation}

We remark that this solution is of the form of that found from the
application of the Lie symmetry vector $X_{3}$, where $a_{1}=\frac{1}{1-K}$
and $u_{0}$ is an integration constant.

\subsubsection{$\left\{ X_{5}\right\} $}

For case (C) of Proposition 1 and the vector field $X_{5}$ we calculate the
similarity transformation $u=a_{2}\ln t+U\left( \sigma \right) ,~\sigma =%
\frac{x}{t}$. Hence, by substituting into equation (\ref{sz.04}) \ we find
the trivial similarity solution%
\begin{equation*}
u\left( t,x\right) =a_{2}\ln \left( U_{0}x\right) ,
\end{equation*}%
which is a static solution; and $U_{0}$ is an integration constant.

\subsubsection{$\left\{ X_{6}\right\} $}

Finally, for case (D) of Proposition 1 and the vector field $X_{6}$ we
derive the similarity transformation $u=-\frac{1}{K}\ln t+U\left( \sigma
\right) ,~\sigma =\frac{x}{t}\,\ $and the closed-form solution%
\begin{equation*}
u\left( t,x\right) =-\frac{1}{K}\ln \left( U_{0}x\right) .
\end{equation*}%
The latter is a static solution, similar to that determined by the vector
field $X_{5}$, when $a_{2}=-\frac{1}{K}$ and $U_{0}$ is an integration
constant.

\subsection{Equation $gZE^{2}$}

We continue our analysis with the study of the similarity transformations
for the 2+1 partial differential equation $gZE^{2}$. At this point it is
important to mention that in order to reduce equation $gZE^{2}$ to an
ordinary differential equation we should make use of two elements of the
one-dimensional optimal system.

However, when we apply the similarity transformation of a given symmetry to
the differential equation, the reduced equation does not admit always the
remainder of the symmetries. Hence, if $Y_{i},~Y_{k}$ are two symmetry
vectors of $gZE^{2}$, with commutator $\left[ Y_{i},Y_{k}\right] =cY_{k}$%
~where $c$ may be zero, application of the invariants of $Y_{i}$ gives a
reduced equation for which $Y_{k}$ is not a point symmetry. Specifically $%
Y_{k}$ becomes a nonlocal symmetry \cite{leach1}. On the other hand,
reduction with respect to the vector field $Y_{k}$ on the master equation
provides a reduced equation which possesses the symmetry vector $Y_{i}$~\cite%
{Govinder2001}$\,$.

Hence, in order to perform a double reduction we should determine all the
two-dimensional subalgebras which consist of the elements of the
one-dimensional optimal system of equation $gZE^{2}$. \ With the use of
Table \ref{tab1} we find the following two-dimensional subalgebras%
\begin{equation*}
\mathbf{A}^{1}=\left\{ \alpha _{1}Y_{1}+\alpha _{2}Y_{2}+\alpha _{3},\beta
_{1}Y_{1}+\beta _{2}Y_{2}+\beta _{3}Y_{3}\right\} ~,~\mathbf{A}^{2}=\left\{
\alpha _{1}Y_{1}+\alpha _{2}Y_{2},Y_{4}\right\} ~,\mathbf{~A}^{3}=\left\{
Y_{3},Y_{4}\right\} ~,
\end{equation*}%
\begin{equation*}
\mathbf{A}^{4}=\left\{ Y_{4},Y_{A}\right\} ~,~\mathbf{A}^{5}=\left\{ \alpha
_{1}Y_{1}+\alpha _{2}Y_{2}+\alpha _{A}Y_{A},\beta _{1}Y_{1}+\beta
_{2}Y_{2}+\beta _{A}Y_{A}\right\} ~.
\end{equation*}

\subsubsection{Reduction with $\mathbf{A}^{1}$}

Application of the similarity transformation provided by the two-dimensional
Lie algebra$\mathbf{~A}^{1}$ in equation (\ref{sz.05}) gives the travelling
wave similarity transformation 
\begin{equation}
u=u\left( \xi \right) ,~\xi =\frac{y\left( a_{1}b_{2}-b_{1}a_{2}\right)
+x\left( b_{1}a_{3}-a_{1}b_{3}\right) +t\left( a_{2}b_{3}-a_{3}b_{2}\right) 
}{a_{1}b_{2}-a_{2}b_{1}}.
\end{equation}

Thus, the reduced equation is determined to be 
\begin{eqnarray}
0 &=&\left( \left( a_{2}b_{3}-a_{3}b_{2}\right) ^{2}-\left(
a_{3}b_{1}-a_{1}b_{3}\right) ^{2}\right) \left( a_{1}b_{3}-b_{1}a_{3}\right)
\left( \frac{u_{\xi \xi }}{f\left( u\right) }\right) _{\xi \xi }  \notag \\
&&+2\left( a_{2}b_{3}-a_{3}b_{2}\right) \left( a_{1}b_{3}-b_{1}a_{3}\right)
\left( g\left( u\right) \right) _{\xi \xi }.
\end{eqnarray}%
Consequently, the reduced equation is similar to that for the generalized
1+1 Zoomeron equation found above, i.e. equation (\ref{fd1}).

\subsubsection{Reduction with $\mathbf{A}^{2}$}

From the Lie algebra $\mathbf{A}^{2}$ we calculate the similarity
transformation%
\begin{equation}
u=u\left( \zeta \right) ~,~\zeta =\left( a_{2}t-a_{1}x\right) y.
\end{equation}

In the new variables the reduced equation is%
\begin{equation}
\left( a_{2}-a_{1}\right) \left( \frac{\left( \zeta u_{\zeta }\right)
_{\zeta }}{f\left( u\right) }\right) _{\zeta \zeta }+2\left( g\left(
u\right) \right) _{\zeta \zeta }=0.
\end{equation}%
Therefore for $\left( a_{2}-a_{1}\right) \neq 0$ it follows that%
\begin{equation}
\left( \zeta u_{\zeta }\right) _{\zeta }+2\left( g\left( u\right) \right)
f\left( u\right) +\left( u_{1}\zeta +u_{0}\right) f\left( u\right) =0.
\end{equation}

We assume now that $f\left( u\right) $ and $g\left( u\right) $ have the
functional forms of that of some specific cases of Proposition 2. For Case
(I), we determine the closed-form solution 
\begin{equation}
u\left( \zeta \right) =u_{0}\zeta ^{a_{1}}.
\end{equation}

Similarly for Case (III) we derive the closed-form solution%
\begin{equation}
u\left( \zeta \right) =u_{0}\ln \zeta ,u_{0}=-\frac{a_{2}}{a_{2}K+2}.
\end{equation}

\subsubsection{Reduction with $\mathbf{A}^{3}$}

From the Lie algebra $\mathbf{A}^{3}$ we find that $u=u\left( \frac{x}{t}%
\right) $. Thus the analytic solution is 
\begin{equation*}
g\left( u\right) =u_{1}\frac{x}{t}+u_{0}.
\end{equation*}

\subsubsection{Reduction with $\mathbf{A}^{4}$}

We continue with the application of the elements of the two-dimensional Lie
algebra $\mathbf{A}^{4}$. \ For simplicity of our presentation we give the
similarity solution and the corresponding exact solution that we can derive.

\paragraph{Case (I)}

The similarity transformation is $u=U\left( \sigma \right) \left( ty\right)
^{a_{1}},~\sigma =\frac{x}{t}$.\ A closed-form exact solution is derived to
be~%
\begin{equation*}
U\left( \sigma \right) =U_{0}\sigma ^{a_{1}}~.
\end{equation*}

\paragraph{Case (II)}

The similarity transformation is $u=U\left( \sigma \right) \left( ty\right)
^{\frac{1}{1-K}},~\sigma =\frac{x}{t}$.\ A closed-form exact solution is
derived to be%
\begin{equation*}
~U\left( \sigma \right) =U_{0}\sigma ^{\frac{1}{1-K}}~.
\end{equation*}

\paragraph{Case (III)}

The similarity transformation is $u=U\left( \sigma \right) +a_{2}\ln \left(
ty\right) ,~\sigma =\frac{x}{t}$.\ A closed-form exact solution is derived
to be~%
\begin{equation*}
U\left( \sigma \right) =a_{2}\ln \sigma .
\end{equation*}

\paragraph{Case (IV)}

The similarity transformation is $u=U\left( \sigma \right) +\frac{1}{1-K}\ln
\left( ty\right) ,~\sigma =\frac{x}{t}$.\ A closed-form exact solution is
derived to be~%
\begin{equation*}
U\left( \sigma \right) =\frac{1}{1-K}\ln \sigma .
\end{equation*}

We omit the presentation with reduction with respect to the two-dimensional
Lie algebra $\mathbf{A}^{5}$ because we were not able to find any
closed-form solution.

\section{Conclusion}

\label{sec5}

In this study we apply the theory of Lie symmetries in order to study the
1+1 and 2+1 generalized Zoomeron equations with two arbitrary functions. We
found that for nonconstant functions, the generalized 1+1 Zoomeron equation
admits two or three Lie point symmetries which form the $2A_{1}$ or the $%
A_{3,3}$ Lie algebras respectively. Moreover, the generalized 2+1 Zoomeron
equation admits four or five Lie point symmetries for specific forms of the
unknown functions.\ The Lie symmetries form the $A_{4.5}^{ab}$ or $%
3A_{1}\otimes _{s}2A_{1}$ Lie algebras respectively. It is important to
mention that the maximum dimensional Lie algebra corresponds, in both cases,
to the non-generalized Zoomeron equations.

For each admitted Lie algebra we determined the one-dimensional optimal
system. The elements of the latter system were used to define similarity
transformations in order to simplify the studied equations.\ Finally, new
exact solutions were found. The latter solutions are singular and can be
seen as the asymptotic solutions which describe the kink behaviour of the
solitons.

In a future work we plan to investigate the existence of analytic solutions
for the reduced systems as also their integrability properties. Such an
analysis overpasses the purposes of this work.

\bigskip

\textbf{Conflicts of interest/Competing interests:} The authors declare no
conflict of interest.

\textbf{Availability of data and material:} Not applicable.

\end{document}